\documentclass[aps,pra,preprint,showpacs,showkeys]{revtex4}
\usepackage{amsmath,amssymb}  
\usepackage{rotating,epsfig} 
\usepackage{bm}  
\usepackage{color}    
\begin{document}

\title{
Dissociative ionization of the H${_2}$O molecule induced by medium energy singly charged projectiles 
}

\author{S. T. S. Kov\'acs}
\email{kovacs.sandor@atomki.mta.hu} 
\author{P. Herczku}
\author{Z. Juh\'asz}
\author{L. Sarkadi}
\author{L. Guly\'as}
\author{B. Sulik}
\affiliation{Institute for Nuclear Research, 
Hungarian Academy of Sciences (MTA Atomki), P.O. Box 51, H-4001 Debrecen, Hungary}

\date{\today}

\begin{abstract}

We report on the fragmentation of the water molecule by $1$ MeV H$^{+}$, He$^{+}$ and 650 keV N$^{+}$ ion impact. The fragment-ion energy spectra were measured by an electrostatic spectrometer at different observation angles. The obtained double-differential fragmentation cross sections for N$^{+}$ is found to be more than an order of magnitude higher, than that for H$^{+}$. The relative ratios of the fragmentation channels are also different for the three projectiles. Additional fragmentation channels were observed in the spectra for He$^{+}$ and for N$^{+}$ impact, which are missing in the case of H$^{+}$. From the analysis of the kinetic energy of the fragments, the maximum observed degree of ionization was found to be $q\rm{_{max}}=3$, $4$, and $5$ for H${^+}$, He${^+}$ and N${^+}$ impact, respectively. Absolute multiple ionization cross sections have been determined. They are compared with the predictions of the classical trajectory Monte Carlo and continuum-distorted-wave eikonal-initial-state theories. At lower degrees of ionization, theories provide reasonable agreement with experiment. The systematic overestimation of the cross section by the theories towards higher degrees of ionization indicates the failure of the independent particle model.

\end{abstract}

\pacs{34.50.Gb, 34.10.+x}
\keywords{molecular collisions, molecular fragmentation, ionization, multiple ionization, CTMC, CDW-EIS}

\maketitle

\section{Introduction}

The dissociation of small few-atomic molecules has been extensively studied by the impact of various types of projectiles, such as photons \cite{Pian01, Ped02, Duj03}, electrons \cite{deBarr04, Frem05}, protons \cite{Luna06, Luna07, Itzhak08}, and multiply charged ions \cite{Gracia09, Alvarado10, Werner11, Rajput12, Sob13}. Molecular fragmentation by ion impact is a rather complex process, which is highly interesting in different areas from astrophysics to cancer therapy. In these fields the most interesting impact energy region is the surrounding of the so called Bragg peak, where the energy transfer to the medium maximizes \cite{Mont14, Luna15, Scha16}. The equilibrium charge state of the projectile ions in the distal region of the Bragg peak is usually close to unity in a wide kinetic energy range, e.g., heavier ions are strongly screened there \cite{Paul17}. In spite of their relevance, systematic studies in the Bragg-peak region with dressed-ion projectiles are rather scarce. In one of those works Montenegro {\em et al.} \cite{Mont14} found that the fragmentation yield does not follow the steep decrease of the linear energy transfer (LET) at the low-energy side of the Bragg peak. The dissociation yield has been found practically constant down to very low projectile energies.  

The fragmentation pattern of a target molecule is determined by the velocity, charge state and structure of the projectile \cite{Alvarado10} and the open fragmentation channels taking place in the reaction \cite{Gracia09}. By the collision, the target molecule may fall to several possible excited and ionized states. Some of those states of the transient (=precursor) molecular ion will initiate dissociation into the open fragmentation channels. Multiple vacancy states are particularly dissociative. Multiple electron removal from the target molecule can happen e.g., by direct multiple ionization, by transfer-ionization or by single ionization followed by secondary processes. Scully \cite{Scully18} {\em et al.} showed that the role of secondary Auger-processes is non-negligible in producing multiply charged molecular ions even in the case of electron impact. The emitted fragments can be neutral or charged particles in excited or ground states \cite{Luna07}. There are also two or three-step fragmentation processes \cite{Luna07, Alvarado10, Itzhak19}, i.e. sequential dissociations such as H${_2}$O${^{2+}}$ $\rightarrow$ OH${^+}$+H${^+}$ $\rightarrow$ O${^+}$+H${^+}$+H${^0}$. Note that a fragmentation channel is usually characterized by the charge states of the precursor ion and the fragments, without specifying their electronic, vibrational or rotational states. Accordingly, the same channel notation may be used for a set of sub-channels with different kinetic energy release (KER) values.  

The kinetic energy release is typically low for ion-neutral breakups. For few-atomic molecules, it is often only a few tenths of eV, and the upper limit is around $5$ eV. For breakups involving at least two positive ions the region of KER extends up to much higher energies, it is between cca. $3$ and $100$ eV \cite{Itzhak19}. The higher KER for the latter case is due to the Coulomb repulsion between the charged fragments, which increases with the charge state of the transient molecular ion (Coulomb explosion) \cite{Werner11, Werner20}. For the accurate determination of the KER distribution one has to take into account the electronic excitation of the transient molecular ion and the emergent fragments \cite{Werner20, Tari21}, as well as the rotational and vibrational degrees of freedom of the precursor molecular ion \cite {Werner11, Rajput12}. The several possible excited states of the precursor molecular ion and the emergent fragments result in a spread of the kinetic energies of the fragments originating from the same dissociation channel. Furthermore, the kinetic energy distribution for a certain fragmentation channel may differ in the case of one-, two- or three-step processes \cite{Luna07, Itzhak19}. As a result, the fragment energy spectra are rather complex.

In most of the experiments, the dissociation pattern of water was studied thoroughly only for the low charge state transient molecular ion (H${_2}$O$^{q+}$, where $q\leqslant3$ )\cite{deBarr04, Frem05, Gracia09, Alvarado10, Mont14, Luna15, Sered22, Sob23}. In a recent work Pedersen {\em et al.} \cite{Ped02} studied the dissociation of the H${_2}$O${^{2+}}$ molecular ion in details, induced by XUV photons from  H${_2}$O${^{+}}$ ions. They devoted special attention to the excited states of the initial molecular ion and the emitted fragments, and their effect on the KER distribution. Higher ionization states of the water molecule were observed in collisions with slow, highly charged ions (HCI) \cite {Rajput12,Sob13,Pesic24,Pesic25,Olivera26}, where the dominant ionization process is multiple electron capture. Here the degree of target ionization strongly depends on the initial charge of the projectile: The maximal degree of target ionization was found to be $q=4, 5,$ and $8$ by different groups utilizing Ne$^{7+}$ \cite{Sob13}, Ar$^{9+}$ \cite{Rajput12} and Xe$^{44+}$ \cite{Olivera26}, respectively. Recently Wolff and co-workers \cite{Wolf36} observed higher degrees of ionization of the water molecule ($q=4, 5$) by the impact of MeV energy ions. 

For heavier ions only relatively few works \cite{Mont14,Wolf36} cover both the charge state and energy ranges, which are typical for the close surrounding and the distal region of the Bragg peak.
In the present work we concentrate on this relevant but less investigated area. We study the emission of fragments from the multiple ionization of water, while bombarding it with medium-energy, single charged atomic-ion projectiles. These projectiles mostly interact with the target molecule by weak, screened Coulomb potential, therefore direct single ionization is the dominant process. Classical and quantum mechanical calculations confirm that in such collision systems, the electron emission spectrum is dominated by electrons from single ionization \cite{Kovacs27}. However, in close collisions, the perturbation strength for "dressed" projectiles may approach that for bare projectiles. This is due to the  rapidly decreasing screening effect of the projectile electrons towards smaller impact parameters. In such close collisions the effective charge exceeds the ionic charge for a short time period \cite{Akos28}, and the target feels strong perturbation. Such collision events can produce remarkable double, and multiple ionization yields even for neutral atom impact \cite{Sarkadi29}. Though their contributions may remain low compared with single ionization, they are responsible for the production of the majority of the fragments. The connection between the primary ionization and the subsequent molecular fragmentation has been subject of numerous studies for lower degrees of ionization \cite{Luna07, Alvarado10, Olivera26}. As the degree of ionization becomes higher with increasing perturbation, several new fragmentation channels open. Thus, fragmentation measurements offer a sensitive method for studying multiple ionization of molecules.

In this work we measured double differential fragment-ion emission spectra for the gas phase H${_2}$O molecule by the impact of H${^+}$, He${^+}$ and N${^+}$ ions. From the spectra we determined absolute cross sections for the individual fragmentation channels. The latter procedure is based on extensive earlier studies performed by several research groups \cite{Ped02, Alvarado10, Werner11, Rajput12, Sob13, Luna15, Sered22, Sob23, Pesic24, Pesic25, Olivera26}, in which the overwhelming majority of the fragmentation channels have been identified and their KER data have been determined, dominantly for H$^+$, He$^{q+}$ and HCI projectiles. From the cross sections determined for the individual fragmentation channels we deduced the multiple ionization cross sections for the target molecule. The experimental results are analysed by comparing them with the predictions of the continuum-distorted-wave eikonal-initial-state (CDW-EIS) and the classical trajectory Monre Carlo (CTMC) theories.

\section{Experiment}

The fragmentation of the H${_2}$O molecule was investigated in a standard crossed beam experiment in Atomki, Debrecen \cite{Kovacs27}. Beams of H${^+}$, He${^+}$ and N${^+}$ were provided by a $5$ MV VdG accelerator with energies $1$ MeV/u, $250$ keV/u and $46$ keV/u respectively.

The ion beams were guided through a 15${^\circ}$ deflector chamber in order to keep the charge state of the ions well defined. After the deflector chamber two pairs of electrostatic steerers were mounted in the beamline, as fine-tuning elements. Collimation of the ion beam was performed by a four-jawed slit placed $120$ cm distance from the entrance of the experimental chamber, and a somewhat larger aperture between the four-jawed slit and the chamber. During beam tuning a precisely aligned additional aperture was temporarily  placed just after the entrance of the experimental chamber. This aperture was removed during the measurements. The beam current was measured by a two-staged differential Faraday-cup. A double-layer magnetic shielding reduced the magnetic field to a few mG in the scattering chamber. 

A jet of H${_2}$O vapour was led into the experimental chamber through a $1$ mm diameter nozzle. A pressure regulator with an automatically operated needle valve ensured constant buffer pressure and continuous gas flow regulation. The container of the pre-purified, carbon-free liquid water was attached to the entrance of this pressure regulating system. Dissolved gases were carefully pumped out. The target gas density in the collision volume was 2$\times$10${^{13}}$ cm${^{-3}}$.  The continuous background pressure was around 9$\times$10${^{-7}}$ mbar and 1$\times$10${^{-5}}$ mbar without and with target gas inlet respectively.

The cylindrical scattering chamber of 1000 mm diameter was equipped with rotatable rings. Charged fragments ejected from the collisions were energy analysed by a single stage energy dispersive electrostatic spectrometer fixed on one of the rings. The experimental geometry allowed us to measure the angular distribution of the fragments from $20{^\circ}$ to $160{^\circ}$ relative to the incident ion beam. In order to avoid recombinations caused by the background gases, we used a small, compact spectrometer, close to the collision region. The pass length from the collision center to the channeltron detector was less than 10 cm. The base energy resolution of the spectrometer was 3\%. 

Fragment ion energy spectra at different observation angles were taken from $0.4$ to $200$ eV. Absolute double-differential fragmentation cross sections were obtained by a standard normalization procedure. The effective target length and target gas density have been evaluated by the procedure given in refs. \cite{Kovacs27, Lattouf30}. The transmission of the spectrometer was determined from its geometrical parameters. The efficiency of the channeltron detector ($\eta=0.85 \pm 0.08$) was taken from the literature \cite{Effic31}. 

The statistical error was estimated less than $20$\% for H${^+}$ impact, and far below $10$\% for He${^+}$ and N${^+}$ projectiles in the main, structured region of the spectra (typically in the $3-15$ eV, $3-30$ eV and $3-50$ eV energy range for proton, helium and nitrogen ion impact, respectively). The systematic error was estimated around $25$\% in these energy regions, mostly due to the uncertainty of the detection efficiency. Thus the overall accuracy of the cross section data in the structured region is $\leq$ $30$\%. Below $3$ eV we estimate the systematic error somewhat higher (cca. $40$\%) due to the charging of the oxidized surfaces of the spectrometer. Therefore, the overall accuracy goes up to $40-50$\% here. At higher energies, near the end of the spectra, the overall uncertainty also increases due to the increasing statistical error.

\section{Theoretical considerations}

In a previous work \cite{Kovacs27} we studied the present collision systems by measuring and analysing the energy spectra of the emitted electrons. There the electron emission cross sections were compared with the results of CDW-EIS and CTMC calculations, extended to treat molecular orbitals and screened potentials for describing the electron emission from molecules impacted by dressed projectiles. The details of the theories can be found in \mbox{Refs. \cite{Kovacs27, Sarkadi32, Gulyas33, Gulyas34}}. The models were applied at the level of the independent particle approximation. In the present work we use the same models to describe multiple ionization of the H$_2$O molecule, leading to molecule fragmentation. For the treatment of the multiple vacancy production in the framework of the independent particle model (IPM), the impact parameter formulation is used. For a specific molecular orbital (MO) the calculations yield impact-parameter dependent single-electron probabilities for ionization, $p_{i}(b)$ and electron capture, $p_{c}(b)$. We note that for molecules, the impact parameter is a vector in the plane, which is perpendicular to the projectile trajectory. Moreover, the probabilities are not only impact-parameter, but also orientation dependent. 

The multiple vacancy production, when $n$ electrons are ejected, and $m$ are captured to the projectile from the initial number of electrons $N$ on a specific MO is given by the following multinomial expression:

\begin{equation}
P_{i^n c^m}={N \choose n}{N-n \choose m}p_i^n\,p_c^m\,(1-p_i-p_c)^{N-(n+m)}
\label{eqno2}
\end{equation}
\vskip 0.5cm

For a molecule having $Q$ MOs, the probability of multiple vacancy creation is a product of the contributions of the individual MOs. The probability of creating the ($n_{1}, n_{2},...,n_{Q}; m_{1}, m_{2},...,m_{Q}$) vacancy configuration is given by  

\begin{equation}
P_{i^{n_1,n_2,\,\, ...\,\, n_Q} c^{m_1,m_2,\,\, ...\,\, m_Q}} =   
\prod_{k=1}^Q {N_k \choose n_k} {N_k-n_k \choose m_k} \times p_{ik}^{n_k}\, p_{ck}^{m_k} \,(1-p_{ik}-p_{ck})^{N_k-(n_k+m_k)}
\label{eqno3}
\end{equation}
\vskip 0.5cm

\noindent
where $k=1,...,Q$; $N_{k}$ is the number of the electrons on the $k$th MO; $p_{ik}$, and $p_{ck}$ are the ionization, and capture probabilities from the $k$th MO, respectively; $n_{k}$ is the number of ejected and $m_{k}$ is the number of captured electrons from orbital $k$.

\section{Results and discussions}

\begin{figure}
\includegraphics[width=10cm,angle=0]{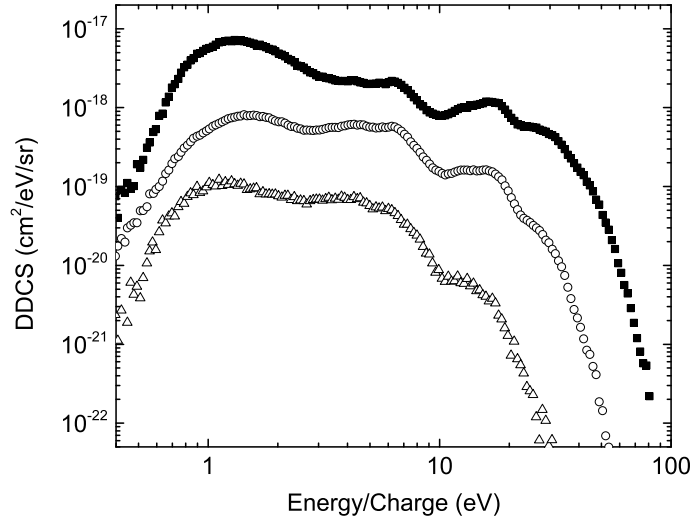}
\vspace*{-0.3cm}
\caption{Absolute double differential fragmentation cross section spectra for H$_2$O molecule measured at $45{^\circ}$ observation angle. Open triangles stand for H$^{+}$ impact, open circles for He$^{+}$, and full squares for N$^{+}$ projectile. The enhanced high energy tail above $15$ eV, which appears for He$^+$ and N$^+$ projectiles, is due to multiple ionization $(q>3)$ processes.}
\label{figure1}
\end{figure}

The fragment ion energy spectra measured in the present work (see Fig. 1) exhibit significant differences for the three projectiles. The fragment ion emission is found to be isotropic, except a high energy tail of the spectra around $90^\circ$ observation angle, which is due to binary collisions between the projectile and one of the target nuclei. Therefore, in Fig. \ref{figure1} we show spectra taken at just one particular observation angle ($45{^\circ}$). The cross section increases with the atomic number of the projectile. It is two orders of magnitude higher for N${^+}$ than that for H${^+}$ impact at all energies. The structure of the spectra also changes significantly with the atomic number of the projectile. 

For the identification of the measured fragmentation channels and fragment energies we leant on the KERs and individual fragment energies given in Refs. \cite{Ped02, Alvarado10, Werner11, Rajput12, Sob13, Rich35}. These values are summarised in Table \ref{tab:table1}. In these studies the fragmentation pattern of the H${_2}$O molecule was investigated by the impact of different projectiles. Pedersen {\em et al.} \cite{Ped02} studied the fragmentation of the H${_2}$O${^{2+}}$ molecular ions by XUV photons in coincident measurements. Special attention was put on the different excited states of the precursor molecular ions and fragments, and their effect on the measured spectra. Alvarado {\em et al.} \cite{Alvarado10} studied the fragmentation of water molecules induced by singly charged ion bombardment. Using time-of-flight technique, they measured the energy distribution of the fragments originating from the single, double and triple ionization of the molecule. Fragments from the higher ionization states of the H${_2}$O molecule were observed in collisions of water with slow HCIs by different groups \cite{Werner11, Rajput12, Sob13}. In a recent work Wolff {\em et al.} \cite{Wolf36} studied the fragmentation pattern of water by different ion projectiles (H${^+}$, Li${^{0...3+}}$, C${^+}$ and C${^{2+}}$). Their fragmentation channel identification was based on a combination of Coulomb explosion and CTMC calculations (CE-CTMC) in reasonable agreement with their experimental results.

We detected only one of the fragments from each dissociation events. However, the fragmentation channels could be well identified in the measured spectra using the information found in the above-mentioned works. From the tabulated KER values in Table \ref{tab:table1}, one can estimate the kinetic energy of the individual fragments for ion-pair breakups by taking into account that the kinetic energies are inversely proportional to the mass of the fragments. Assuming that the neutrals carry a negligible amount of KER \cite{Itzhak19}, this estimation can be extended to ion-pair + neutral breakups, too.

\begin{figure}
\includegraphics[width=10cm,angle=0]{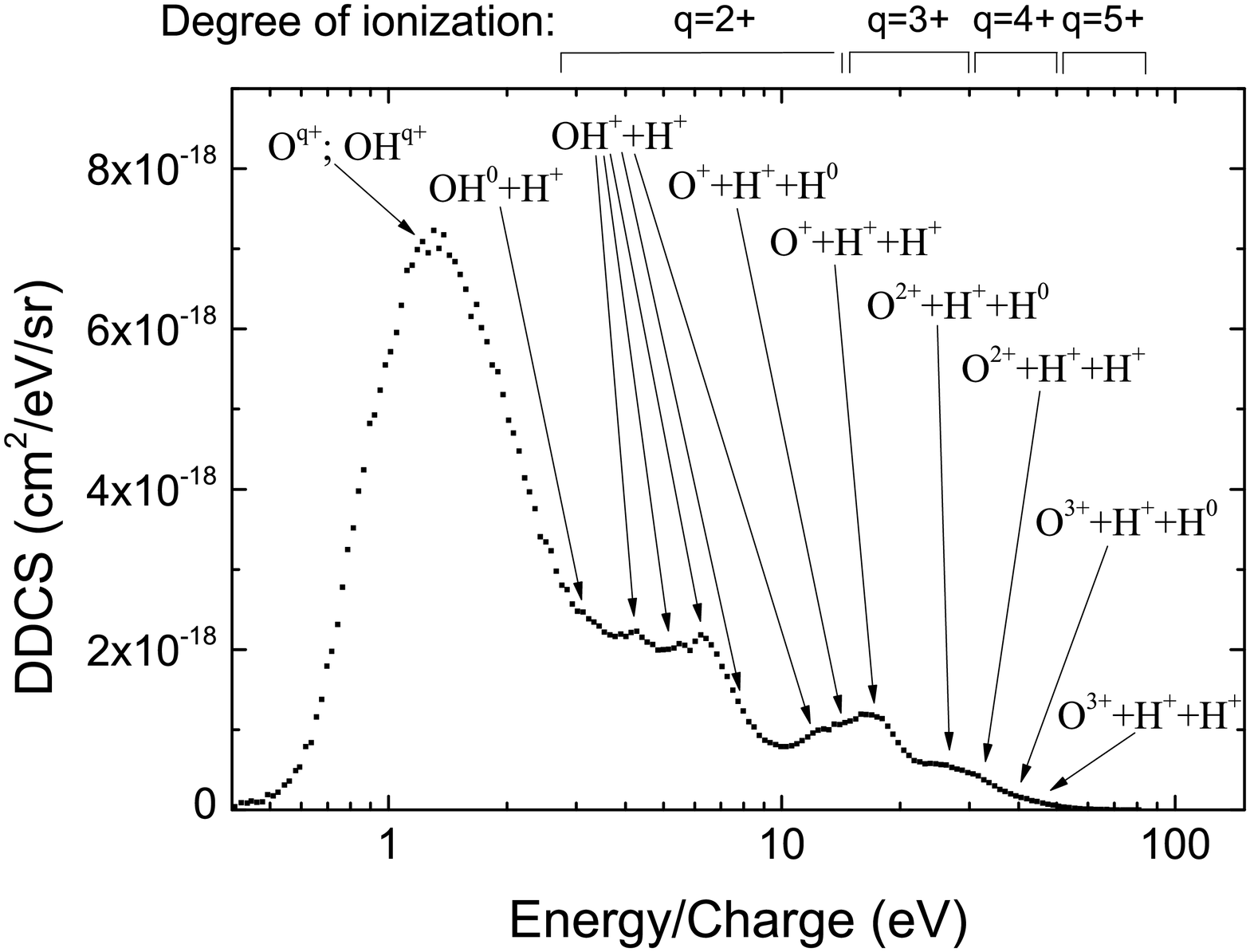}
\vspace*{-0.3cm}
\caption{Absolute double-differential fragmentation cross section spectra of the H$_2$O molecule induced by $650$ keV N$^+$ impact. The presented spectrum was measured at $45{^\circ}$ observation angle. The identified fragmentation channels and the regions of the different ionization degrees are indicated.}
\label{figure2}
\end{figure}

The identified fragmentation channels are shown in Fig. \ref{figure2}. The unresolved hump below $3$ eV reflects mostly heavy fragments (O${^{q+}}$; OH${^{q+}}$) from ion-pair and ion-triplet breakups. A small amount of low energy H${^+}$ ions from ion-neutral breakups (single ionization of water) may also contribute to this region. In the case of ion-pair and ion-triplet breakups proton fragments produce a structured region above ca. $3$ eV. According to Fig. \ref{figure2}, the double ionized water molecule dissociates mostly into two fragments. Protons form the OH${^+}$+H${^+}$ channel produce an almost flat region from ca. $3$ eV to $7$ eV, and a more structured part between $7$ and $12$ eV. It contains several overlapping peaks, which belong to different excitation states of the transient H${_2}$O${^{2+}}$ molecular ion and the emergent OH${^{+}}$ fragments. Similar conclusions were drawn for the overlapping peaks in Refs. \cite{Ped02, Alvarado10, Werner11}. We note here that Refs. \cite{Ped02, Rajput12, Sob13} predict a slight contribution of ion-pair + neutral channels to this energy region. The three-, four- and five-fold ionized molecules dominantly dissociate into ion-triplets \cite{Alvarado10}. Protons from these highly ionized ($q>2$) transient molecular ions appear above $15$ eV. 

\begin{sidewaystable}
\footnotesize
\caption{\label{tab:table1} Summary of the literature data used in the present analysis. The last column refers to the identification number of the peak, which stands for the fragmentation channel in Fig. 3 and Table II. The same number with different lowercase letters represent the components of one "collector" peak during the fit.}

\begin{ruledtabular}
\begin{tabular}{cccccccccc}

Projectile&
Method&
Fragmentation channel &
KER (eV)&
FWHM (eV)&    
H$^+$ energy\, (eV) & 
H$^+$ FWHM\, (eV) & 
Ref. No. &
Peak No. & \\
 
6-23 keV H${^+}$; He${^+}$; He${^{2+}}$
& TOF
& $\rm{H_{2}O^{+} \rightarrow OH^{0}+H^{+}}$ 
        & $-$
        & $-$
        & $2\pm 0.5$ 
        & $-$
        & \cite{Alvarado10}
        & $4$ &\\
      
6-23 keV H${^+}$; XUV 
& TOF; coinc
& $\rm{H_{2}O^{2+} \rightarrow OH^{+}+H^{+}}$ 
        & $3.7\pm 0.5$
        & $-$
        & $3.5\pm 0.5$
        & $-$    
        & \cite{Ped02, Alvarado10}
        & $5$ &\\

6-23 keV He${^+}$; XUV
& TOF; coinc
& $\rm{H_{2}O^{2+} \rightarrow OH^{+}+H^{+}}$
        & $4.0\pm 0.5$
        & $-$
        & $3.8\pm 0.5$
        & $\sim 6$ \cite{Alvarado10}
        & \cite{Ped02, Alvarado10}
         & $6a$ &\\

He II
& PIPICO
& $\rm{H_{2}O^{2+} \rightarrow OH^{+}+H^{+}}$
        & $4.5\pm 0.5$
        & $-$
        & $\sim 4.0$
        & $-$
        & \cite{Rich35}
         & $6b$ &\\  
        
XUV,  He II 
& ion spect
& $\rm{H_{2}O^{2+} \rightarrow O^{+}+H^{+}+H^{0}}$ 
        & $4.8\pm 1.0$
        & $-$
        & $~5$
        & $-$
        & \cite{Sob13, Rich35}
         & $6c$ &\\        

1-5 keV He${^{2+}}$
& TOF; coinc
& $\rm{H_{2}O^{2+} \rightarrow OH^{+}+H^{+}}$
 		& $6.8\pm 1$
 		& $-$
 		& $6.5\pm 1$
 		& $\sim 6$
 		& \cite{Alvarado10}
 		 & $7a$ &\\

XUV
& coinc
& $\rm{H_{2}O^{2+} \rightarrow OH^{+}+H^{+}}$
 		& $6\pm 2$
 		& $-$
 		& $-$
 		& $-$
 		& \cite{Ped02}
 		 & $7b$ &\\
        
XUV
& coinc
& $\rm{H_{2}O^{2+} \rightarrow OH^{+}+H^{+}}$
 		& $7.8$
 		& $-$
 		& $-$
 		& $-$
 		& \cite{Ped02}
 		 & $8$ &\\
 		
XUV
& coinc
& $\rm{H_{2}O^{2+} \rightarrow OH^{+}+H^{+}}$
 		& $11.7$
 		& $-$
 		& $-$
 		& $-$
 		& \cite{Ped02}
 		 & $9$ &\\
        
6-23 keV H${^+}$; He${^+}$
& TOF
& $\rm{H_{2}O^{2+} \rightarrow O^{+}+H^{+}+H^{0}}$ 
        & $15.3$
        & $-$
        & $14.5\pm 1$
        & $\sim 15$
        & \cite{Alvarado10}
         & $10$ &\\

6-23 keV He${^{2+}}$
& TOF
& $\rm{H_{2}O^{3+} \rightarrow O^{+}+H^{+}+H^{+}}$ 
        & $36$
        & $-$
        & $17.8\pm 0.45$
        & $\sim 15$
        & \cite{Alvarado10}
         & $11a$ &\\

20 keV HCI
& TOF, ion spect
& $\rm{H_{2}O^{3+} \rightarrow O^{+}+H^{+}+H^{+}}$ 
        & $\sim 35$
        & $-$
        & $18\pm 1.0$
        & $\sim 15$
        & \cite{Werner11, Rajput12, Sob13}
         & $11b$ &\\
      
6-23 keV He${^{2+}}$
& TOF
& $\rm{H_{2}O^{3+} \rightarrow O^{2+}+H^{+}+H^{0}}$ 
        & $\sim 30$
        & $-$
        & $28\pm 0.5$
        & $\sim 23$
        & \cite{Alvarado10}
         & $13a$ &\\

20 keV HCI
& ion spect
& $\rm{H_{2}O^{3+} \rightarrow O^{2+}+H^{+}+H^{0}}$ 
        & $-$
        & $-$
        & $28\pm 1$
        & $-$
        & \cite{Rajput12, Sob13}
         & $13b$ &\\    
        
100-125 keV HCI
& TOF
& $\rm{H_{2}O^{4+} \rightarrow O^{2+}+H^{+}+H^{+}}$ 
        & $\sim 68$
        & $\sim 20$
        & $-$
        & $-$
        & \cite{Werner11}
         & $14$ &\\

20 keV HCI
& ion spect
& $\rm{H_{2}O^{4+} \rightarrow O^{3+}+H^{+}+H^{0}}$ 
        & $-$
        & $-$
        & $38\pm 2$
        & $-$
        & \cite{Sob13}
         & $15$ &\\

100-125 keV HCI
& TOF
& $\rm{H_{2}O^{5+} \rightarrow O^{3+}+H^{+}+H^{+}}$ 
        & $\sim 95$
        & $\sim 28$
        & $-$
        & $-$
        & \cite{Werner11}
         & $16$ &\\

\end{tabular}
\end{ruledtabular}
\end{sidewaystable}

A further analysis of the spectra in Fig. 1 revealed that the proton fragment peaks above ca. 15 eV, appearing only for He${^+}$ and N${^+}$ projectiles, belong to the fragmentation channels which are due to the four- and five-fold ionization of the water molecule. In parallel with these proton peaks, the relative yield of the heavy fragments ($<3$ eV) also increases, which can be understood considering that light fragments have their corresponding heavy partners. From the maximum kinetic energy of the emitted protons, we concluded that the highest degree of ionization was $q\rm{_{max}}=3$, 4, and 5 for H${^+}$, He${^+}$ and N${^+}$ impact respectively.

The fragmentation spectrum for N${^+}$ impact observed in the present work is very similar to those reported in Refs. \cite{Sob13, Pesic25} measured by slow highly charged ions  (see Fig. 3b in Ref. \cite{Sob13}). At first sight the perturbation exerted by the single charged nitrogen projectile seems to be surprisingly strong. The strong multiple ionization capability of the dressed N${^+}$ projectile can be attributed to the reduced screening of the projectile nucleus by its electrons in close collisions,  i.e., at small N-O impact parameters, where multiple ionization is dominant. Accordingly, the effective charge for multiple ionization may exceed the ionic charge significantly.

\begin{figure}
\includegraphics[width=10cm,angle=0]{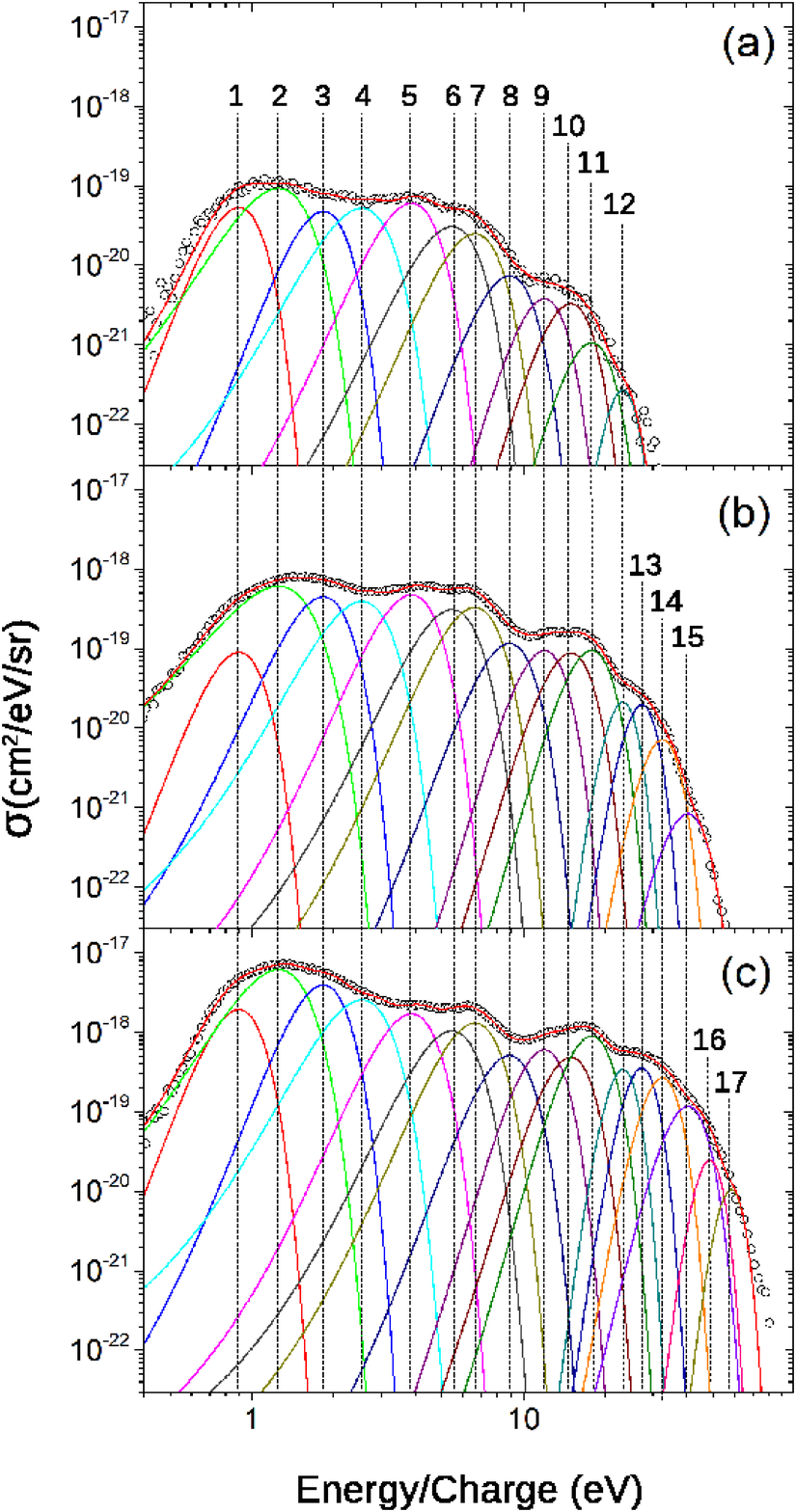}
\vspace*{-0.3cm}
\caption{Fragment ion spectra of H$_{2}$O induced by H$^{+}$ (a), He$^{+}$ (b), and N$^{+}$ (c) projectiles (symbols). The peaks represent Gaussian fit for the fragmentation channels listed in Table 1 and Table 2. Channel positions are indicated by vertical lines with numbers.}
\label{figure3}
\end{figure}

For a quantitative analysis of the measured fragmentation patterns the spectra were decomposed to contributions from particular fragmentation channels. The fit curves together with the measured data are shown in Fig. \ref{figure3}. The fit is based on the data listed in Table \ref{tab:table1} and on our channel identification (Fig. \ref{figure2}). The region of heavy fragments ($<3$ eV) is covered by 3 Gaussians. Their mean energies and FWHMs were varied to achieve the best fit. In the region of protons, each Gaussian represents an identified fragmentation channel. Only one peak around $23.2$ eV was inserted for the best fit. The energy center of the Gaussians was kept fixed during the fit. For the N$^{+}$ projectile an additional peak around $58$ eV was necessary to insert at the end of the spectrum. This peak is likely to be due to fragments from the highly excited, five-fold ionized H$_{2}$O$^{5+}$* molecule. Data about the FWHM values are scarce in the literature. The few FWHM values for the individual channels presented in Table \ref{tab:table1} have large uncertainties. Nevertheless, for charged particle impact, these FWHM values are roughly proportional to the mean channel energies. Therefore, to resolve the problem of the incomplete knowledge of the FWHM data, we assumed that the widths of the peaks are proportional to their energy centers. The initial value of the proportionality factor was set to 0.4. In an iterative fitting procedure this factor was allowed to vary in a range between 0.2 and 0.4. In a small extent we allowed also slight changes (max. 5\%) of the peak centers. Finally the same set of the FWHM and energy center values were used to fit the spectra for all the three projectiles. The results of the fit are shown in Fig. \ref{figure3} and in Table \ref{tab:table2}.

\begin{sidewaystable}
\footnotesize
\caption{\label{tab:table2} The obtained cross sections ($\sigma$) of the individual fragmentation channels for the three projectiles. They are the results of the fit presented in Fig. \ref{figure3}. The energy center and FWHM values are the same for all projectiles. The uncertainties includes only the statistical errors and the estimated uncertainties of the fit.}

\begin{ruledtabular}
\begin{tabular}{cccccccc}

Peak No.&
Fragmentation channel &     
 Center \, (eV) & 
 FWHM \, (eV) & 
 $\rm{\sigma_{{H^+}} \, (cm^{2})}$ & 
 $\rm{\sigma_{{He^+}} \, (cm^{2})}$ &
 $\rm{\sigma_{{N^+}} \, (cm^{2})}$ &\\

		$1$
		& $\rm Heavy \,\, (OH^{q+}$; $\rm O^{q+})$ 
        & $0.89$
        & $0.35$
        & $2.54\pm 0.15\times10{^{-19}}$
        & $4.33\pm 0.41\times10{^{-19}}$
        & $9.16\pm 0.31\times10{^{-18}}$ &\\

		$2$
		& $\rm Heavy \,\, (OH^{q+}$; $\rm O^{q+})$
        & $1.25$
        & $0.65$
        & $8.18\pm 0.25\times10{^{-19}}$
        & $6.19\pm 0.08\times10{^{-18}}$
        & $5.39\pm 0.06\times10{^{-17}}$ &\\
        
		$3$
		& $\rm Heavy \,\, (OH^{q+}$; $\rm O^{q+})$
        & $1.83$
        & $0.74$
        & $4.75\pm 0.24\times10{^{-19}}$
        & $4.79\pm 0.08\times10{^{-18}}$
        & $3.87\pm 0.06\times10{^{-17}}$ &\\
        
		$4$
		& $\rm{H_{2}O^{+} \rightarrow OH^{0}+H^{+}}$
        & $2.54$      
        & $1.23$
        & $8.66\pm 0.27\times10{^{-19}}$
        & $6.49\pm 0.08\times10{^{-18}}$
        & $4.21\pm 0.05\times10{^{-17}}$ &\\

		$5$
		& $\rm{H_{2}O^{2+} \rightarrow OH^{+}+H^{+}}$
        & $3.87$
        & $1.68$
        & $1.36\pm 0.03\times$10${^{-18}}$
        & $1.07\pm 0.01\times$10${^{-17}}$
        & $3.82\pm 0.05\times$10${^{-17}}$ &\\

		$6$
		& $\rm{H_{2}O^{2+} \rightarrow OH^{+}+H^{+}}$
		& $5.46$
        & $2.45$
        & $1.02\pm 0.05\times$10${^{-18}}$
        & $1.02\pm 0.02\times$10${^{-17}}$
        & $3.40\pm 0.09\times$10${^{-17}}$ &\\
        
		$7$
		& $\rm{H_{2}O^{2+} \rightarrow OH^{+}+H^{+}}$
		& $6.63$
        & $2.83$
        & $9.51\pm 0.46\times$10${^{-19}}$
        & $1.25\pm 0.02\times$10${^{-17}}$
        & $4.93\pm 0.10\times$10${^{-17}}$ &\\
        
		$8$
		& $\rm{H_{2}O^{2+} \rightarrow OH^{+}+H^{+}}$
		& $8.89$
        & $3.50$
        & $3.46\pm 0.23\times$10${^{-19}}$
        & $5.43\pm 0.10\times$10${^{-18}}$
        & $2.40\pm 0.07\times$10${^{-17}}$ &\\
  
		$9$
		& $\rm{H_{2}O^{2+} \rightarrow OH^{+}+H^{+}}$
		& $11.94$
        & $4.22$
        & $2.15\pm 0.19\times$10${^{-19}}$
        & $5.35\pm 0.10\times$10${^{-18}}$
        & $3.40\pm 0.08\times$10${^{-17}}$ &\\

		$10$
		& $\rm{H_{2}O^{2+} \rightarrow O^{+}+H^{+}+H^{0}}$
		& $14.93$ 
        & $5.33$
        & $2.37\pm 0.20\times$10${^{-19}}$ 
        & $6.27\pm 0.14\times$10${^{-18}}$ 
        & $3.47\pm 0.11\times$10${^{-17}}$ &\\

		$11$
		& $\rm{H_{2}O^{3+} \rightarrow O^{+}+H^{+}+H^{+}}$
		& $17.83$
        & $6.11$
        & $8.76\pm 1.23\times$10${^{-20}}$
        & $7.77\pm 0.11\times$10${^{-18}}$
        & $7.34\pm 0.10\times$10${^{-17}}$ &\\

		$12$
		& $\rm{H_{2}O^{3+} \rightarrow ?}$
		& $23.24$
        & $5.30$
        & $1.91\pm 0.34\times$10${^{-21}}$
        & $1.51\pm 0.05\times$10${^{-18}}$
        & $2.42\pm 0.06\times$10${^{-17}}$ &\\

		$13$
		& $\rm{H_{2}O^{3+} \rightarrow O^{2+}+H^{+}+H^{0}}$
		& $27.30$
        & $6.59$
        & $9.14\pm 4.39\times$10${^{-22}}$
        & $1.72\pm 0.04\times$10${^{-18}}$
        & $3.14\pm 0.06\times$10${^{-17}}$ &\\

		$14$
		& $\rm{H_{2}O^{4+} \rightarrow O^{2+}+H^{+}+H^{+}}$ 
		& $32.52$ 
        & $8.82$
        & $-$ 
        & $8.53\pm 0.28\times$10${^{-19}}$
        & $3.17\pm 0.06\times$10${^{-17}}$ &\\

		$15$
		& $\rm{H_{2}O^{4+} \rightarrow O^{3+}+H^{+}+H^{0}}$ 
		& $40.36$ 
        & $12.75$
        & $-$ 
        & $1.43\pm 0.11\times$10${^{-19}}$ 
        & $1.99\pm 0.05\times$10${^{-17}}$ &\\

		$16$
		& $\rm{H_{2}O^{5+} \rightarrow O^{3+}+H^{+}+H^{+}}$ 
		& $48.58$ 
        & $10.05$
        & $-$ 
        & $-$ 
        & $3.32\pm 0.22\times$10${^{-18}}$ &\\
        
		$17$
		& $\rm{H_{2}O^{5+} \rightarrow O^{4+}+H^{+}+H^{0}}$ 
		& $58.22$ 
        & $11.67$
        & $-$ 
        & $-$ 
        & $1.65\pm 0.09\times$10${^{-18}}$ &\\

\end{tabular}
\end{ruledtabular}
\end{sidewaystable}

We note that the fitted curves in the $4-12$ eV energy region may contain slight contributions of fragmentation channels different from the identified components of the OH$^+$+H$^+$ channel. According to the published experimental data, the $\rm{H_{2}O^{2+} \rightarrow O^{+}+H^{+}+H^{0}}$ \cite {Rajput12, Sob13, Rich35} channel also provides a small yield between 5 and 6 eV. Calculated data in Ref. \cite{Ped02} suggest that the $\rm{H_{2}O^{2+} \rightarrow O^{0}+H^{+}+H^{+}}$ fragmentation channel may also contribute to the 4$-$12 eV region, but it has not been detected in any experimental work. As the articles report only small yields for these channels, the questioned energy region is attributed to the OH$^+$+H$^+$ fragmentation channel in our work, characterized by slightly different KER values (see Table \ref{tab:table1}).

The mean energies of the fragmentation channels of three-, four- and five-fold ionization fall above 17 eV. They agree well with those of calculated by the nCTMC model of Wolff {\em et al.} \cite{Wolf36}. Their channels denoted by {\em e-h} in Ref. \cite{Wolf36} can be identified with our channels No. 13-16 in Table II, respectively. At lower energies the number of peaks in Ref. \cite{Wolf36} is significantly smaller, though they can be identified with some of the peaks found in the present fittings. The reason is that many of the considered channels, taken from the literature, belong to excited states of the ionized precursor molecule, while no excitation is included in the model of Wolff {\em et al.} \cite{Wolf36}. Nevertheless, their predicted energy positions are surprisingly good for the ground state of the precursor molecule ion.

The analysis of the obtained cross sections, presented in Table \ref{tab:table2} shows that the highest energy fragmentation channels have almost two orders of magnitude lower yield than the double ionization channels for all projectiles. The highest-energy proton fragments belong to the fragmentation channels of $\rm O^{2+}+H^{+}+H^{0}$, $\rm O^{3+}+H^{+}+H^{0}$ and $\rm O^{4+}+H^{+}+H^{0}$ for H$^+$, He$^+$ and N$^+$ impact respectively. 

\begin{figure}
\includegraphics[width=10cm,angle=0]{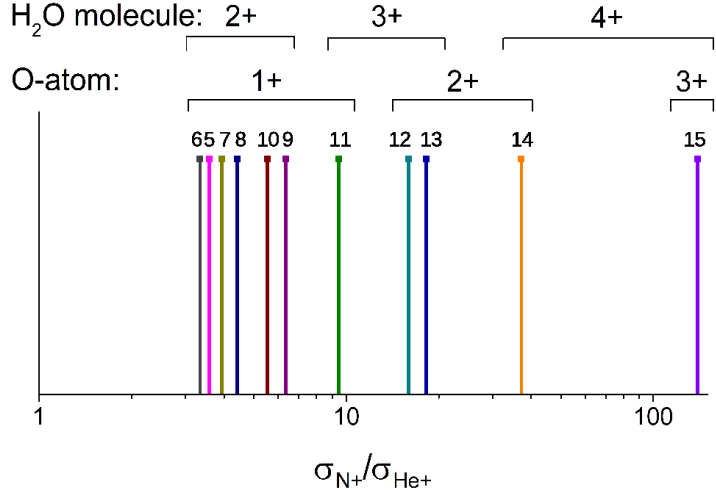}
\vspace*{-0.3cm}
\caption{The $\sigma_{\rm N^+} / \sigma_{\rm He^+}$ ratios for the individual fragmentation channels (the colours of the lines are the same as the colours of the Gaussians in Fig. \ref{figure3}). The ratios are presented only from the OH$^+$+H$^+$ channel. The sequence of the lines is almost the same as the sequence of the energy centers of the Gaussians. The ionization degree of the H$_2$O molecule and the O-fragments are indicated.}
\label{figure4}
\end{figure}

Further analysis was made via the $\sigma_{\rm N^+} / \sigma_{\rm He^+}$ ratios of the individual fragmentation channels (Fig. \ref{figure4}). It is expected that this ratio is increasing with higher degrees of ionization. Indeed, the ratios form groups according to the degree of ionization of the molecule, and subgroups according to the degree of ionization of the oxygen atom. As expected, the ratio is an almost monotonic function of the energy of the proton fragment. It is seen that the multiple ionization efficiency of the N$^+$ projectile relative to that of He$^+$ dramatically increases with the degree of ionization.

From the results of the fit we deduced the multiple ionization cross sections of the water molecule as sums of the partial cross sections of the corresponding individual fragmentation channels. Single ionization cross sections could not been determined with the present method. The main reason is that the non-dissociative single ionization events can not be detected by our method at all. Another reason is that in the $0.4-3$ eV energy region, due to the strong overlap of the peaks, it is not possible to separate the heavy fragments from the light H$^+$ fragments originating from single ionization. Moreover the kinetic energy of some of the fragments from ion-neutral breakups falls below our detection limit ($0.4$ eV). 

Double ionization of the H${_2}$O molecule may easily happen by removing both electrons from one of the  O-H bonds. Accordingly, there is a rather large probability that one of the chemical bonds breaks, while the other remains unharmed. This can be the reason for the relatively large yield of the OH${^+}$+H${^+}$ channel. For higher degrees of ionization both O-H bonds are likely to be affected. Therefore, the probability of ion-pair breakups becomes negligible, and the molecule prefer to dissociate into three parts.

The experimentally obtained multiple ionization cross sections (CS) are compared to those calculated by the CTMC and CDW-EIS method. Multiple ionization data produced by the two theoretical methods are also compared with each-other. The detailed description of the models is given in Refs. \cite{Kovacs27, Sarkadi29, Sarkadi32, Gulyas33, Gulyas34}. We found in our previous study \cite{Kovacs27} that CTMC provided good agreement with the measured double differential electron emission cross sections for all the present collision systems. The results of the CDW-EIS calculations also reproduced the experimental double differential electron emission cross sections for H$^+$ and He$^+$ projectile, but they shown significant deviations for N$^+$ impact. In the present work, we concentrate on the total probabilities and cross sections for ionization and electron capture. At this level, both theories predict that electron emission is dominated by single ionization. 

\begin{figure}
\includegraphics[width=10cm,angle=0]{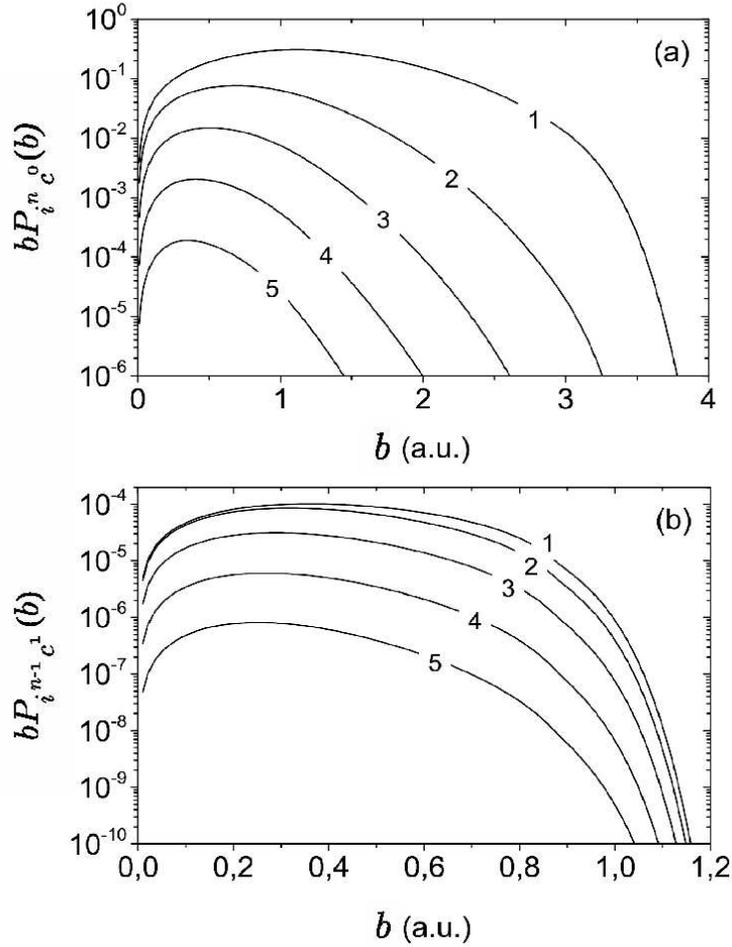}
\vspace*{-0.3cm}
\caption{Ionization (a), and single capture + ionization (b) probabilities as function of the impact parameter for the H$^{+}$ + H$_{2}$O collision system. The number of vacancies produced in the target molecule, ${n}$ is indicated at the curves.}
\label{figure5}
\end{figure}

\begin{figure}
\includegraphics[width=10cm,angle=0]{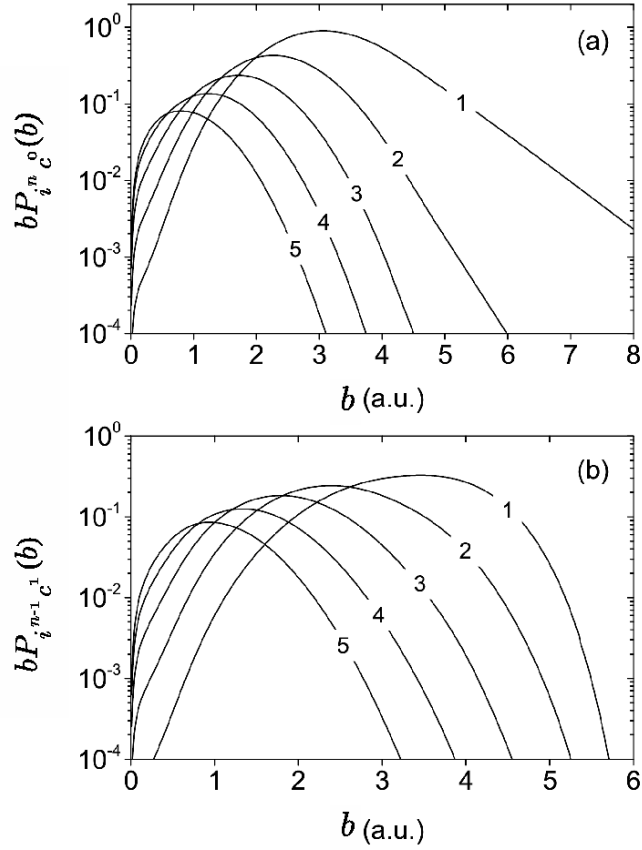}
\vspace*{-0.3cm}
\caption{Ionization (a), and single capture + ionization (b) probabilities as function of the impact parameter for the N$^{+}$ + H$_{2}$O collision system. The number of vacancies produced in the target molecule, ${n}$ is indicated at the curves.}
\label{figure6}
\end{figure}

In the following we analyse the multiple target vacancy production for water predicted by the two theories within the framework of the independent particle model (IPM). For a descriptive presentation we derive orientation-averaged $P_{i^n c^m}(b)$ values (see Eq. (1)) for $n$-fold ionization and simultaneous $m$-fold electron capture as a function of a scalar impact parameter  $b$. This way, we can demonstrate and compare the approximate impact parameter dependence of the multiple vacancy creation probabilities. In Fig. \ref{figure5}, we present CTMC results for $n$-fold ionization (a) and for singe-electron capture + $(n-1)$-fold ionization (b). The averaged $bP_{i^n,c^0}(b)$ and $bP_{i^{n-1} c^1}(b)$ curves for H$^{+}$ impact on H$_{2}$O are plotted in Fig. \ref{figure5}a and Fig. \ref{figure5}b respectively. The impact parameter dependence of the same processes for N$^{+}$ impact is shown in Fig. \ref{figure6}.  In the figures, the impact parameter is "measured" from the nucleus of the oxygen atom. Note that the areas under the $bP(b)$ curves are proportional to the cross sections of the particular processes. 

According to the CTMC results for H$^{+}$ impact, single ionization is dominant in the full impact parameter region. The yields of higher degrees of ionization become more significant in narrower regions of smaller impact parameters (see Fig. \ref{figure5}a). They remain much below the single ionization yield everywhere. The maximum of the calculated $bP(b)$ curves decreases about three orders of magnitude from single- to five-fold ionization. Single capture + ionization is limited to a small impact parameter range, and its contribution to vacancy production is negligible at all degrees of ionization. The shape of the $bP_{i^{n-1} c^1}(b)$ curves for different $n$-s are very similar to each other. (See Fig. \ref{figure5}b). 

The relevant impact parameter region for ionization is much larger for N$^{+}$ than that for H$^{+}$ projectile (See Fig. \ref{figure6}a). Single ionization is also dominant here in the whole $1-8$ $a.u.$ impact parameter region with a maximum around $3$ $a.u.$. The multiple ionization curves for $N^+$ impact extend to impact parameter ranges that are twice as large as those for proton impact. Similarly to H$^+$ impact, increasing degrees of ionization have smaller yields in gradually narrower windows at smaller impact parameters. However, the decrease of the yields is much weaker here: the maximum of the curve is only about one order of magnitude smaller for five-fold than for single ionization. In contrary to H$^{+}$ impact, multiple ionization curves exceed that for singe ionization at small impact parameters (below 2 $a.u.$). It shows that the effective perturbation strength increases towards smaller impact parameters. Moreover, it indicates that this is due to the screened potential of a $Z=7$ central charge, which goes far above the ionic potential at small distances. This behaviour is even more pronounced for the single capture + ionization process, as it is seen in Fig. \ref{figure6}b.

\begin{figure}
\includegraphics[width=10cm,angle=0]{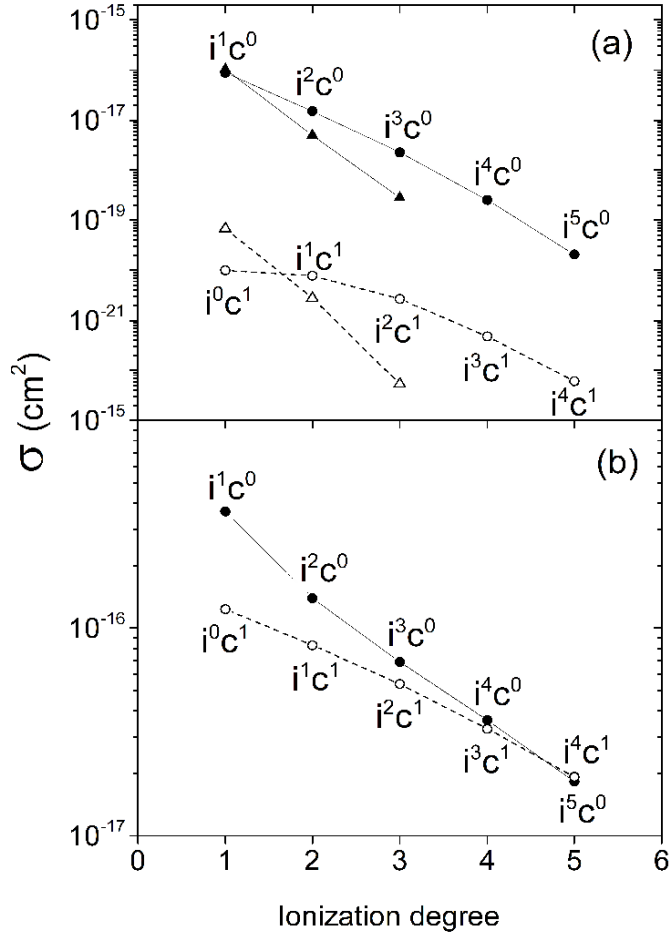}
\vspace*{-0.3cm}
\caption{Pure ionization (i$^n$ c$^0$) and single capture + ionization (i$^{n-1}$ c$^1$) cross sections for H$^{+}$ (a), and for N$^{+}$ (b) projectiles. Pure ionization is presented as full circles for CTMC and full triangles for CDW-EIS calculations. Capture + ionization process is presented by open circles and open triangles for CTMC and CDW-EIS respectively. The lines are for guide the eye.}
\label{figure7}
\end{figure} 

The calculated multiple vacancy production cross sections for H$^+$ and N$^+$ impact are shown in Fig \ref{figure7}. According to the CTMC calculations the target ionization cross section for H$^+$ impact decreases more than three orders of magnitude from single to five-fold ionization. Cross sections calculated by CDW-EIS for single, double, and triple ionization are also presented in Fig \ref{figure7}a. They decrease faster with increasing degree of ionization than those obtained by the CTMC method. According to both theories the electron capture contribution to the vacancy production is negligible for the H$^+$ projectile. The yield of single-electron capture + ionization events remains at least two orders of magnitude lower than that of pure ionization leading to the same number of vacancies. 

For N$^+$ impact, the absolute cross sections are significantly larger, and their relative yields are strongly different from those of H$^+$ impact. The decrease of the cross section with increasing number of vacancies is much slower here, only one order of magnitude from single to five-fold ionization. Moreover, the role of electron capture is not negligible for N$^+$ impact. With increasing degree of target ionization the cross sections of the two processes approach each other. The cross section for single-capture + four-fold ionization even exceeds that of pure five-fold ionization (See Fig. \ref{figure7}b).  

\begin{figure}
\includegraphics[width=10cm,angle=0]{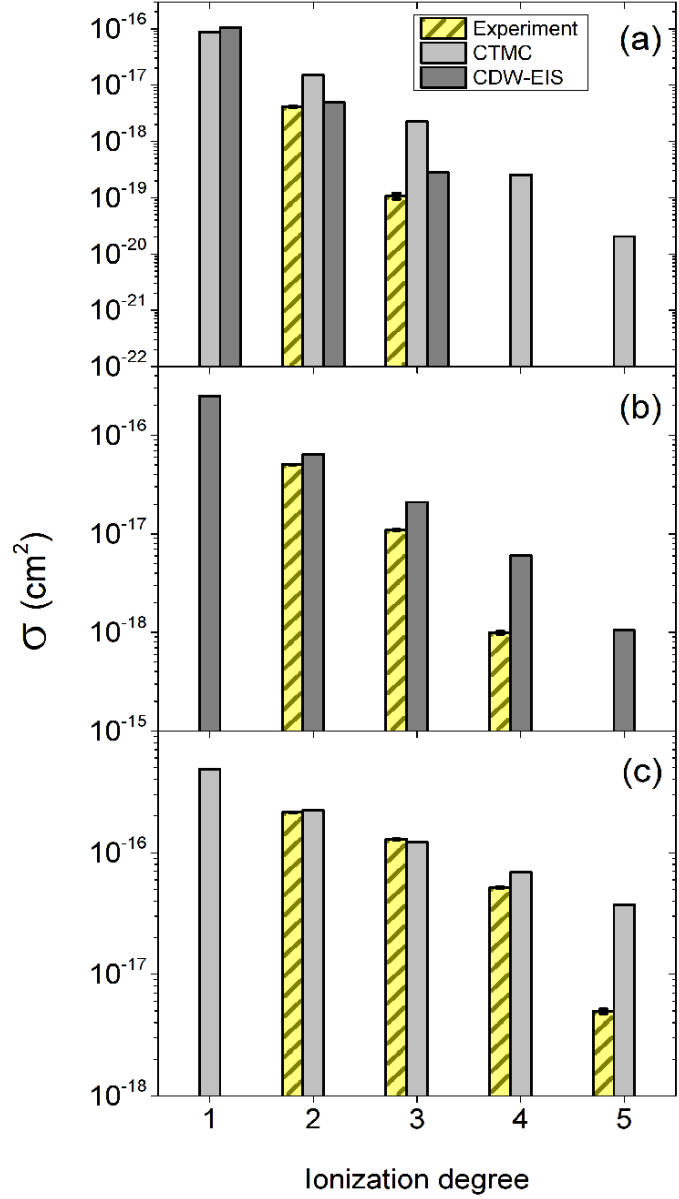}
\vspace*{-0.3cm}
\caption{Multiple ionization cross section as function of the ionization degree of the target for H${^+}$ (a), He${^+}$ (b) and N${^+}$ (c) bombardment. The theoretical predictions for the different ionization degrees are also shown.}
\label{figure8}
\end{figure}

In Figure \ref{figure8}, the experimentally determined multiple ionization cross sections are compared with those obtained by CDW-EIS and CTMC calculations. For double vacancy production, CDW-EIS provides a good agreement with experiment for both H$^{+}$ and He $^{+}$ impact at $1$ MeV projectile energy. Moreover, there is a reasonably good agreement with CDW-EIS for the triple vacancy yields too. This quantum treatment seems to perform better than CTMC at high impact velocities and small perturbations, as it is seen for H$^{+}$ impact. 

We could not measure single ionization in the present experiment directly. Nevertheless, we note that we have experimental information about it. In our earlier work \cite{Kovacs27} we measured the electron emission from the same collision systems, and determined absolute double differential cross sections for it. Those data have been compared with the results of both CDW-EIS an CTMC calculations at the level of double differential spectra. Good agreement was found with CDW-EIS results for H$^{+}$ and He $^{+}$ impact, and with CTMC results for all the three projectiles (H$^{+}$, He $^{+}$ and N$^{+}$).Therefore, we may also consider the theoretical predictions for single ionization as "semi-experimental" values.

For He$^+$ the CDW-EIS results also agree well with the measured multiple target ionization cross sections (See Fig. \ref{figure8} (b)). The agreement is as good as for proton impact in the case of double and triple ionization. However, the experimental four-fold ionization cross section is far below the prediction of the theory. A slight five-fold ionization cross section is also predicted but it was not found in the measurements. For the slowest N$^{+}$ projectile, the CTMC results practically coincide with the experiment up to triple ionization (See Fig. \ref{figure8} (c)). For four-fold ionization, there is a slight deviation. Only the five-fold ionization cross section is overestimated significantly. Since we compare absolute cross sections, this agreement is remarkable. 

A closer inspection of Figure \ref{figure8} shows a general tendency, namely that the measured multiple target ionization cross sections decrease faster with the degree of ionization than the calculated data. It is true for both theoretical models. While for double target ionization both calculations provide reasonable agreement with experiment, they both tend to gradually overestimate the experimental data towards higher ionization degrees. At four-fold and five-fold ionization this tendency becomes very strong. This increasing deviation of the calculated data from the experiment can be attributed to the limitations of IPM. The role of electron correlation in electron emission increases with the degree of ionization. When a single ionization probability is calculated with the first ionization potential as parameter, IPM is expected to overestimate the multiple electron removal from the target. 

Our data show that this overestimation is stronger if the perturbation is weak, and becomes less significant with strong perturbation. While for H$^{+}$, He $^{+}$ projectiles the theories overestimate the cross section for $n=3$, and dramatically overestimate it for $n=4$, for N$^{+}$ impact the agreement is perfect for $n=3$ and still reasonable for $n=4$. It breaks down only at $n=5$.
This finding suggests that the importance of electron correlation may depend on the ratio of a mean correlation energy to a mean energy transfer characteristic for the collision.

\section{Summary and Conclusions}

We studied the fragmentation of H$_{2}$O molecules by the impact of 1 MeV energy H$^{+}$, He$^{+}$, and 0.65 MeV energy N$^{+}$ projectiles. Single charged ions in this energy region are relevant for studying ion $+$ H$_{2}$O collisions in the distal region of the Bragg peak. The energy and angular distribution of the emerging fragments were measured by a single stage, parallel-plate type, electrostatic spectrometer in a standard, crossed beam experiment. Absolute double differential fragmentation cross sections of water were obtained for the three collision systems. The fragment energy spectra were fitted by Gaussian functions, and absolute cross sections for the particular fragmentation channels have been determined. From those channel yields we deduced the multiple ionization cross sections for the water molecule, and compared them with those calculated by CTMC and CDW-EIS methods.

The identification of the particular fragmentation channels is based on their experimental KER values published in the literature.
We found that up to five-fold ionization, the list of the fragmentation channels is close to complete. Moreover, we confirmed that a recent theoretical approach \cite {Wolf36} provided correct identification and reasonable KER values for an important fraction of the fragmentation channels.

We found that the fragment ion emission was isotropic for all projectiles. The differential fragmentation cross section for N$^{+}$ is more than four times larger than that for He$^{+}$, and almost two order of magnitude higher than that of  H$^{+}$ in the entire fragment energy region. This strong variation of the yields is attributed to the increasing perturbation strength of the slower and slower projectile ions from  H$^{+}$ to N$^{+}$. Besides the absolute differences between the cross sections, the relative ratios of the individual fragmentation channels are also different for the three projectiles, and additional channels appear for He$^{+}$ and even more for N$^{+}$ impact towards the high energy end of the spectra. The presence of these fragmentation channels indicate that the maximum ionization degree increases from H$^{+}$ to N$^{+}$. It was found to be $q\rm{_{max}}=3$, $q\rm{_{max}}=4$, and $q\rm{_{max}}=5$ for H${^+}$, He${^+}$ and N${^+}$ impact respectively. 

The fragmentation cross section spectrum for N$^{+}$ impact is very similar to those obtained by slow HCIs. This similarity indicates that the perturbation strength for the N$^{+}$ projectile can approach those for HCIs. This is partially due to the increase of the effective projectile charge in close collisions with the oxygen atom of the target. The dominance of low impact parameter events in the production of multiple ionized H$_{2}$O$^{q+}$ molecular ions ($q=2,... \, 5$), is confirmed by  CTMC calculations.

The experimentally determined absolute multiple ionization cross sections are in a general agreement with the results of the classical CTMC and the quantum mechanical CDW-EIS calculations at lower degrees of ionization. At small perturbations CDW-EIS provides better agreement with experiments than the CTMC model.  For N$^{+}$ impact, the non-perturbative character of the classical CTMC method gains importance. At this strong perturbation, the agreement between CTMC and experiment is remarkably good up to triple, and it remains reasonable even for four-fold ionization. 

Towards higher ionization states both theories systematically more and more overestimate the experimental cross sections. We attribute it to the limitations of the independent particle model, namely the neglect of electron correlation within the IPM framework. In addition, we found that the overestimation is stronger if the perturbation is weak, and becomes less significant with strong perturbation. For N$^{+}$ impact the agreement with experiment holds up to four-fold ionization. This finding suggests that the importance of electron correlation may depend on the ratio of a mean correlation energy to a mean energy transfer characteristic for the collision. 

In conclusion, we studied the distal (i.e., the low energy) part of the Bragg peak in ion - water molecule collisions both experimentally and theoretically. We found that our CDW-EIS and CTMC models are able to provide quantitative account for the multiple ionization of the target molecule in a wide range of the perturbation strength. Quantum calculations (CDW-EIS) proved to be more accurate for weak perturbations, while the non-perturbative CTMC method provided excellent agreement with experiment for violent collisions. We also gained information about the relative importance of electron correlation for weak and strong perturbations. We expect that a combined application of the tested theoretical methods will provide a satisfactory level of quantitative description in this focal region of different applications.

\section{Acknowledgements}

This work has been supported by the Hungarian Scientific Research Foundation (OTKA Grant No.: K109440), and by the National Information Infrastructure Program (NIIF). The authors thank the VdG-5 accelerator stuff for the careful operation.

\end{document}